# Characterizing the transition from classical to quantum as an irreversible loss of physical information

P. Guillemant, C. Abid and M. Medale

Aix-Marseille University, CNRS, IUSTI UMR 7343, 13453 Marseille, France

The experimental confirmation of Landauer's principle and the emerging concept of a computational universe make it more and more crucial to understand the physical sense of information, as it has an intrinsic relation with observer knowledge that is often rejected as subjective. In this paper we propose an objective definition of phase information in a purely classical computational universe with quantized phase states, this quantization being imposed by our fundamental hypothesis that physical information has a finite and limited density, responsible for the irreversibility. We use a statistical study of results obtained by numerical simulations of a billiard to highlight an excessive and paradoxical loss of phase information that we solve by involving a "classical to quantum transition". After discussing the pertinence of such a transition to clarify some problematic aspects of statistical physics, we conclude that a computational universe should automatically lose phase information through time in an irreversible way, which could be compensated by a gain of physical information due to observation and decoherence.

## I. INTRODUCTION

One of the fundamental problems of mainstream physics is the question of how to define physical information, which is very closely linked to entropy and energy. Whereas in quantum physics some part of information provided by a measurement can be observer dependant, this is not the case in classical physics where all phase states are considered to be deterministic, so that their intrinsic information should necessarily be independent of observation. In support of this objective conception, the idea that information is physical - based on the fact that erasing one bit corresponds to the dissipation of *k log(2)* entropy – as proposed early on by Szillard [1] and highlighted by Landauer [2], has recently been confirmed experimentally [3]. But the information theory of Shannon [4] has introduced subjective entropy [5] defined using probabilities and quantifying all types of information, such as that contained in a message. Consequently, the attempts to define physical information as a genuine physical quantity by avoiding the trap of probabilities [6], that inevitably represents a subjective lack of knowledge, have resulted in a lot of confusion amplified by the eternal debate about how to solve the Maxwell's Demon problematic [7] [8]. Today, one widespread opinion is to consider physical information as related to the computational complexity of a system [9], for example by expressing it as the entropy of a cellular automaton [10] that leads to its calculation. This modern concept of information has the advantage of being more objective and of satisfying the initial concept that information has to come in bits.

Since we disagree with this opinion, in this paper we propose a definition of physical information [11] which is relative to states and phases of elementary objects of a system. We focus on phase information of objects like particles, molecules or balls and we define their information as the number of bits necessary to memorize the phase coordinates of objects in a discrete phase space. Note that the idea of a truly discrete space is seriously considered in physics, for example by Roger Penrose who recently argued in this sense by saying that "*it might turn out that a discrete picture [for a computational space-time] is really the correct one*" [9]. As we do not know the inner structure of such hypothetical space, we have chosen for our study a cartesian grid model. Our purpose is to show that whatever the grid precision or space quantum, we cannot avoid a loss of information - associated with increasing entropy - that is independent of space quantum and observer. This would be a genuine loss of information from the universe particularly likely to occur within non linear dispersive systems. From a purely classical point of view, this means that phase states in such systems could become partially indeterminist like in quantum mechanics. However a deterministic framework could be reestablished if a loss of classical information was equivalent to a transition from classical to quantum states, and this is what we want to argue for. This kind of transition has already been observed, for example in some non linear systems in nanotechnologies [12] or with quantum dots [13].

According to Beckenstein [14], we make the fundamental hypothesis that a discrete computational universe has a finite density of classical information everywhere. The transition from classic to quantum behavior would then be caused by a decrease in information density. Consequently the frontier between classical and quantum states of particles would not be clearly defined because this transition would be continuous, as confirmed by a recent experiment [15], thus introducing the possibility that quantum behavior could affect classical macroscopic objects. Some well-known indicators of this behavior can be seen for example in interferences of giant fullerene molecules [16]. Let us also quote the Gibbs paradox [17] that is only clearly solved in the context of quantum statistical mechanics for which the principle of indiscernability of objects is inherent to their quantum nature. But what about classical objects?

In quantum mechanics, indeterminacy is quantitatively characterized by wave function, which is a probabilistic distribution of the measurable values of a given observable. The epistemic idea we are following is that this probabilism has to be considered as a fundamental indeterminism and not as the result of quantum model incompletion. This idea has recently been reaffirmed by M. F. Pusey et al [18] in a paper arguing in favor of the intrinsic reality of the quantum state. Another aspect of quantum indeterminacy is contained within the Heisenberg principle, relating the standard deviations of position $p$ and momentum $q$ as expressed below:

$$\Delta p \Delta q \geq \frac{h}{2\pi} \tag{1}$$

Here too, the principle of indetermination is subjected to various interpretations, the original one arguing that it is the result of an inevitable disturbance of measurements. However, the more fundamental interpretation of this inequality, asserting the intrinsic indeterminism of quantum states, has recently been confirmed by an experimental result of Lee A. Rozema et al that was obtained by weak measurements [19].

According to our hypothesis, we consider that the intrinsic uncertainties $\Delta p$ and $\Delta q$ of any classical or quantum object are variables that result from a fundamental limitation of its phase information. This limitation can be expressed using the maximum uncertainties $\Delta p_{max}$ and $\Delta q_{max}$ that can be deducted from the geometric and energetic characteristics of the system. So we can express the phase information in the form:

$$I_{pq\_det} = Log_2\left(\frac{\Delta p \max \Delta q \max}{\Delta p \Delta q}\right) \tag{2}$$

The ratio between brackets corresponds to the objective number of microstates of the object. Note that there would be no sense counting states under $\Delta p$ and $\Delta q$ because it would quantify a real uncertainty or lack of physical information. In the case of a quantum particle we know the maximum of residual information that the particle can acquire:

$$I_{pq\_ind} = Log_2\left(\frac{\Delta p \Delta q}{h}\right) \qquad (3)$$

If the phase information of any object was unvarying, we could express it in the form:

$$I_{pq} = I_{pq\_det} + I_{pq\_ind} = Log_2\left(\frac{\Delta p \max \Delta q \max}{\Delta h}\right) \qquad (4)$$

Our hypothesis, is then to consider that, for any object, the physical phase information is expressed by $I_{pq\_det}$ instead of $I_{pq}$, giving it an exclusively classical sense. The possibility for $I_{pq\_ind}$, $I_{pq\_det}$ and $I_{pq\_det}$ to vary is equivalent to the possibility for classical objects to lose information, and the aim of our study is to highlight and quantify this loss. In support of this definition of phase information, when we compare (1) et (3) we can interpret the Heisenberg Principle as being a simple condition for phase information to have a maximum, according to our hypothesis:

$$I_{pq\_ind} \geq 0 \Rightarrow I_{pq\_det} \leq Log_2\left(\frac{\Delta p \max \Delta q \max}{\Delta h}\right) \qquad (5)$$

This way of defining physical information can be understood as the equivalence between a classical and a completely informed state, and between a quantum and partially non-informed state, meaning that the transition between classic and quantum states would be progressive, as revealed by recent Nobel prices Serge Haroche and David Wineland. It permits also to clarify the problem of irreversibility as being linked to the variation of physical information, which is a loss during the classical to quantum transition and a gain during the quantum to classical one, with no possibility to recover the same information as before.

## II. MATERIAL AND METHODS

We present a numerical method to calculate the evolution of global entropy $S$ and individual information (2) during the multiple interactions of classical objects such as balls or molecules. First, we have to establish the expression of $S$ from the summations of $I_{pq\_det}$ in our Cartesian coarse-grained simplified space structure with quanta $\varepsilon_p$ and $\varepsilon_q$ for distance and momentum (for which we would have $\varepsilon_p \varepsilon_q = h$ if the space really had an elementary structure of this type).

A discrete quantification of phase states was proposed very early on by Gibbs [16] using the coarse-graining method to explain the monotonous growth of entropy $S$ within a mixture before achieving equilibrium. However, the loss of information ($-\Delta S$) highlighted by this method cannot be considered as a genuine loss from the system as long as the basic equation from which it derives contains subjective probabilities $p_i$ :

$$S = -k \sum_i p_i \ln(p_i) \tag{6}$$

Where $k$ is the Boltzman constant and $p_i$ are the probabilities of microstates $i$ of the system.

For our calculations we have chosen a two-dimensional billiard system containing identical incompressible balls. We calculated all the elastic shocks between billiard balls and with the borders, beginning with random initial conditions where all positions and momentums were perfectly known, meaning $\Delta p = \varepsilon_p$ and $\Delta q = \varepsilon_q$ ( i. e. completely informed phases). Since we were unable to work with realistic values of quanta $\varepsilon_p$ and $\varepsilon_q$, we considered values varying from $2^{-5}$ to $2^{-35}$ fractions of the used resolution to quantify our results, whose positions were visualized in a 4096 x 4096 square-pixels billiard game.

We used an asymptotic analysis to estimate the results for much lower quantum values and also for high values of the number of balls $N_b$. Our calculations were repeated as long as necessary to obtain statistical results that were robust and independent of the initial conditions. At initial conditions where $\Delta p = \varepsilon_p$ and $\Delta q = \varepsilon_q$, all microstates are equiprobable, so we can rewrite (6) in the form:

$$S = -k \ln\left(\left(\frac{\Delta p \max \Delta q \max}{\Delta p \Delta q}\right)^{N_b}\right) = -N_b k \ln(2) I_{pq\_det} \tag{7}$$

In our system we choose $\Delta pmax = 4096$. As for $\Delta qmax$, it is limited by the sum of our initial moments. When we calculate the shocks, $\Delta p$, $\Delta q$ and then $S$ vary as a consequence of a chaotic dispersion - or Lyapunov effect [20] - that shocks produce in the trajectories.

In order to calculate the evolution of $S$ we had to overcome two difficulties. The first is that the variation of $\Delta p$ and $\Delta q$ is not visible in terms of $I_{pq\_det}$ bits because the number of bits that is used to save any data is a constant which for our calculations is equal to 64 bits. The second difficulty, more problematic, is that at each time the real information of any result decreases while maintaining the same number of bits, it accumulates more and more false information, due to the limitation of computer precision that produces inconsistent bits at lower scales that end up corrupting calculations at the higher scales.

To remedy to the first difficulty, we calculated two different trajectories for each ball - as if they were in two distinct billiards - with an infinitesimal difference on their initial positions equal to $\pm\varepsilon_p$, the plus or minus sign being randomly chosen for each ball and axis of coordinates. We approximated $\Delta p$ by the difference of position between the two trajectories at identical times and $\Delta q$ by the difference of velocities at the instant of shocks with the same ball, these instants being slightly different.

To overcome the second difficulty, we worked with values of $\varepsilon_p$ ranging from $2^{-5}$ to $2^{-35}$ by successive divisions by $2^5$, avoiding values between $2^{-40}$ and $2^{-55}$ that could falsify calculations with the effect of inconsistent low-level data after only a few shocks. We estimated that the using of only 20 to 50 bits compared to the maximum 64 permitted (14 bits never being used) could ensure sufficient reliability for our results over an average time of respectively 25 to 10 shocks per ball.

From the calculated $\Delta p(n)$ and $\Delta q(n)$ values, where n is the number of shocks, we can derive from (7) an expression of the global information of the billiard by dividing this expression by the entropy quantum $k\ ln(2)$, changing to logarithms in base 2:

$$I_b = \sum_{n=1}^{N_b} Log_2\left(\frac{\Delta p\max \Delta q\max}{\Delta p(n)\Delta q(n)}\right) = \sum_{n=1}^{N_b} Log_2\left(\frac{\Delta p\max}{\Delta p(n)}\right) + \sum_{n=1}^{N_b} Log_2\left(\frac{\Delta q\max}{\Delta q(n)}\right) \quad (8)$$

Where:

$$\Delta p(n) = \sqrt{(X(n,1) - X(n,2))^2 + (Y(n,1) - Y(n,2))^2} \quad (9)$$

*X (n,i)* and *Y(n,i)* correspond to the coordinates *(X,Y)* of the trajectory *i*=1, 2 of a ball after n shocks on the billiard *i*. The expression of *Δq(n)* is similar to *Δp(n)* : replace positions with velocities.

We noticed that when we follow the two trajectories of the same ball, whose initial position coordinates have a difference that is *ΔX = $\varepsilon_p$* and/or *Δy = $\varepsilon_p$* and with identical initial velocities, the first shock between one ball and another creates a difference of velocity that increases with each subsequent shock. As a result, it is useless to introduce an initial difference of velocity whose effect is less significant than that strictly produced by the difference of positions. We can conclude that it is pointless to use two quanta $\varepsilon_p$ and $\varepsilon_q$ simultaneously, thus we opt for the only consideration of the position quantum and to ignore the effects of the velocity quantum, which leads to disregarding the second term of (8).

For the same reason, it is pointless to use the position quantum to round up to the appropriate precision the results of calculating each shock - in respect to the coarse-grained grid - because the effect of this rounding-up on one trajectory is negligible in comparison to the effect of the initial quantum difference between the two trajectories. So, our coarse-grained calculation is returned in a study which consists of following the evolution of two quasi-superposed billiards, using the maximum precision permitted by the computer. However, a residual difficulty is the necessity of synchronizing the two billiards balls so as to optimize computing time.

As results for each trajectory are highly dependent on initial conditions, we perform a statistical study by averaging the obtained values of *Δp* for many initially randomized trajectories so as to calculate *Δpmoy* and then to compute the following average variation of information:

$$I_b = \sum_{n=1}^{N_b} Log_2\left(\frac{\Delta p \max}{\Delta p(n)}\right) = \max(I_b) - N_b Log_2\left(\frac{\Delta pmoy}{\varepsilon_p}\right) \tag{10}$$

With:

$$\max(I_b) = N_b Log_2\left(\frac{\Delta p \max}{\varepsilon_p}\right) = N_b P_i \tag{11}$$

Where $P_i$ is the bit precision of initial conditions, which varies between 15 and 55 bits. So, our study consisted of calculating the evolution of $\Delta pmoy$ and for this purpose, we chose a time unit equal to the average number of shocks per ball.

In addition to $\varepsilon_p$ we investigate the influence of two other parameters, which are the number of shocks per ball and the void ratio $Rv$ that play an important role in the loss of information, particularly the rate of the loss.

The continuous decay of the information leads to the apparition of an important phase that is occurring when one value of $\Delta p$ becomes large enough locally to cause the divergence of the trajectories histories of the two-paired billiards. This happens more precisely when the shocks of two coupled balls occur with no more coupled ones.

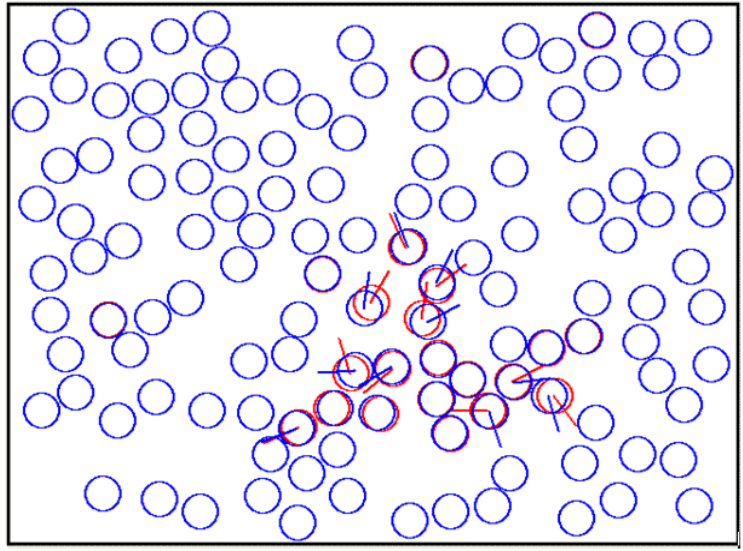

Fig. 1. Illustration of the critical moment of divergence of trajectory histories of two initially superposed billiards with $N_b$=128, $\varepsilon_p$ =$2^{-35}$ and $Rv$=0.33. Calculations were stopped just before history decorrelation. Straight lines show velocity vectors when the distance between red and blue balls exceeds 1 pixel.

After this critical step, the distance separating two coupled balls (the blue and red balls in figure 1) suddenly increases and the two billiard balls are no longer superposed on the screen; this occurs when the difference $\Delta p$ exceeds an average value of 1 pixel. At this stage, our calculations have to be stopped because it is no longer possible to synchronize the two billiard balls. Moreover, it is obvious that after this step, the overall information rapidly decays to zero, for two reasons. First, the velocity vectors of diverging coupled balls become extremely different and second, the resulting contagion effect quickly spreads to the rest of the billiard. Thus we can estimate that after a short delay, shorter than the average time required for a single ball to cross the table, all the balls have completely lost their phase information.

We will henceforth ignore this phase of contagion that follows the critical step by considering that the billiard's history becomes undetermined once the distance between two balls exceeds 1 pixel. For a given quantum $\varepsilon_p$, this critical step is then considered as a transition from classical to quantum states of billiard balls, since after this moment we observe the coexistence of multiple possible histories. So the determination of the average delay before reaching this step has been the main focus of our study. This delay is a dimensionless value that is equal to the average number $Nc$ of shocks per ball prior to the transition.

An example of the variation of $Nc$ with $\varepsilon_p$ is illustrated on Figure 2, where each critical step is the moment when the two trajectories (the blue and the red) begin to diverge.

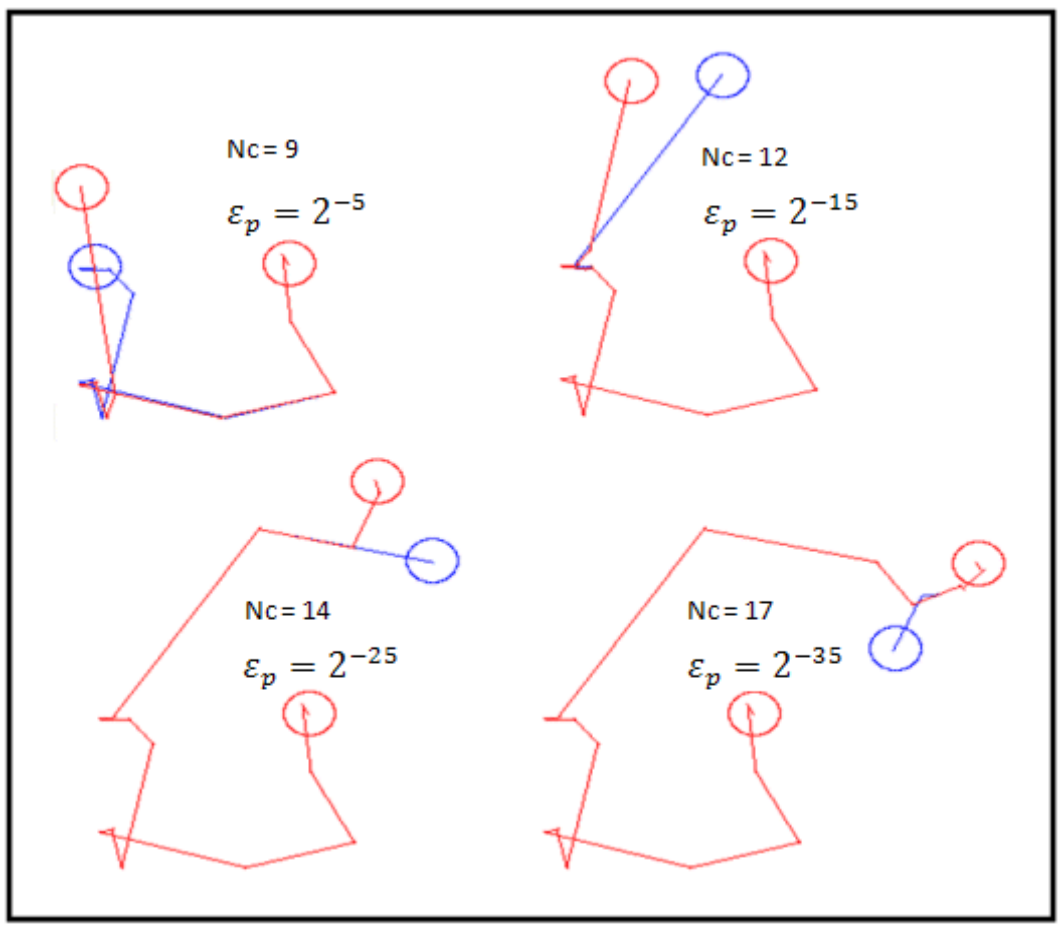

Fig. 2. A computed example of four couples of superimposed ball trajectories. As they originally only differ from between $2^{-5}$ to $2^{-35}$ pixels, the blue one is not visible before reaching the critical step where a divergence or decorrelation occurs. Note that each division of $\varepsilon_p$ by $2^{-5}$ is just delaying this critical step by 2 or 3 shocks.

On Figure 3 we present the results of our statistical study for two methods of calculation, the first (unbroken lines) taking into account the shocks with borders and the second (dotted lines) using a periodic billiard where each ball leaving the game returns to the other side with the same velocity. As the first method has a faster calculation time, we applied it for our global statistic where all values of all parameters were tested. The results presented on Figure 3 have been calculated with a void ratio equal to 0.33 and a number *Nb* of balls ranging from 8 to 512. We observe a good correlation between the two methods of calculation for high values of *Nb*, which is explained by the reduction of the cushion effects. We also observe an approximately linear evolution of Nc versus $Log_2(Nb)$ for both cases which supports the qualitative results of Figure 2, but a very slight decrease of the slope towards high values of *Nb* is also to be noted, testifying to a probable non linear evolution for *Nb* > 512. However, due to the important statistical fluctuations, this result does not allow us to determine a robust asymptotic evolution able to explain the behavior of the curves for higher values of *Nb*. For $\varepsilon_p = 2^{-5}$ the average number of shocks reaches the minimum value of *Nc*=1. This minimum is not zero because when a divergence occurs we stop the calculation and memorize the corresponding number of shocks, which is at least 1, so the *Nc* average cannot be less than 1. In order to calculate an estimation of the critical time we then have to consider the instant *Nc*-1.

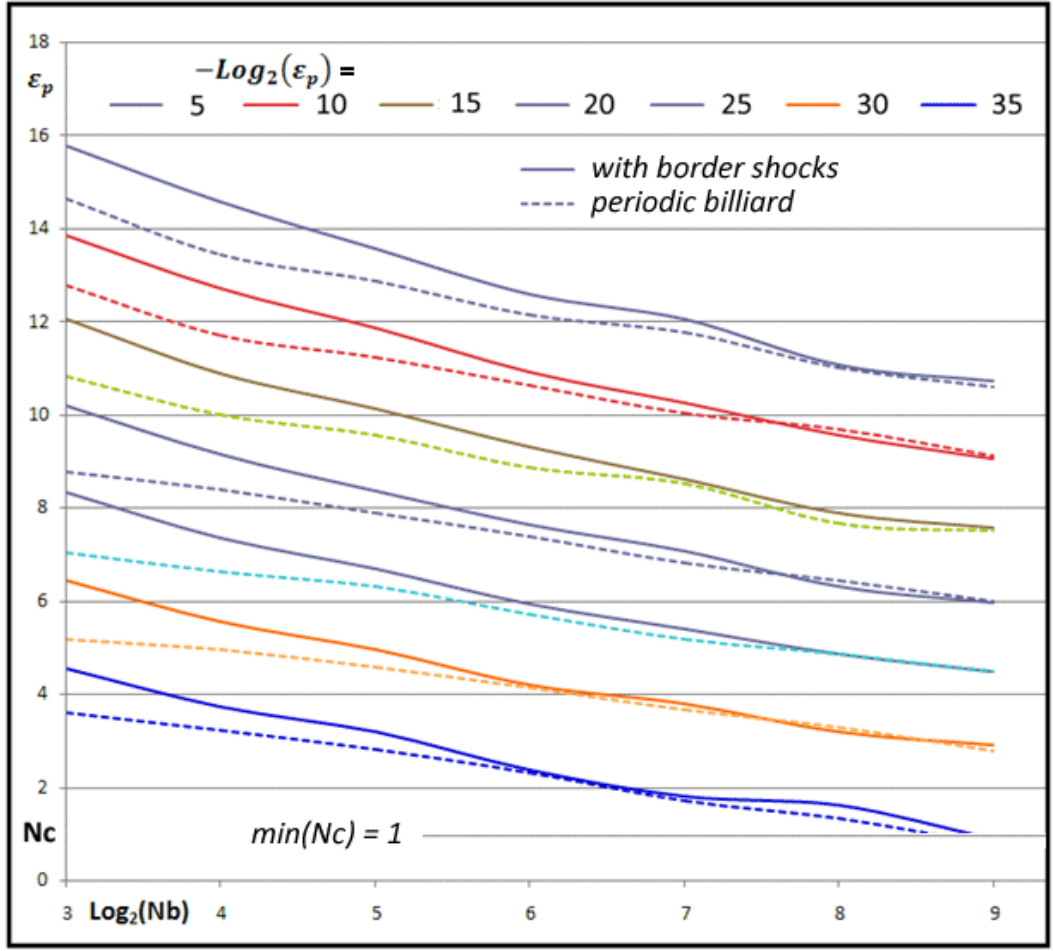

Fig. 3. Results of statistical study with two billiard models, the first one (unbroken lines) by taking into account shocks with the borders and the second one (dotted lines) using a periodic billiard: border effects become negligible for *Nb*>128. We observe an approximately linear evolution of *Nc* (critical instant) with Log2(*Nb*) and Log2($\varepsilon_p$)

To reach the value *Nc*=1, a single shock is sufficient for the whole billiard, provided it is a diverging one. As it is the case for $\varepsilon_p = 2^{-5}$ and *Nb* > 256 (see Fig 3), we can expect that for higher values of $\varepsilon_p$ the possibility exists for other curves to cross the horizontal axis *Nc*=1 when *Nb* is high enough. In order to verify this point we adopted a new calculation method with the aim of greatly reducing computing time by avoiding the calculation of approximately *Nt* x *Nb* x *Nc* shocks, where *Nt* is the number of statistical tests.

This new method consists in a two-balls model where we consider all geometrically possible shocks to calculate the velocity dispersion probabilities. For this purpose we could first calculate the probability for each shock occurring and then consider partial derivatives with respect to angle and position so as to calculate the local dispersion function. As this analytical method was extremely complex we preferred to use our direct simulation of real billiards to calculate the global distribution of velocity dispersions, by using a very large set of randomly initiated calculations corresponding to hundreds of thousands of shocks.

We calculate the dispersion function in the form of the following histogram $A_n$:

$$A_n = 128 + 4 Log_2 \left( \frac{|V_{in1} - V_{in2}|}{|V_{out1} - V_{out2}|} \right) \quad (12)$$

$|V_{in1} - V_{in2}|$ and $|V_{out1} - V_{out2}|$ are the modules of the velocity differences between coupled balls before and after shocks. The 128 and 4 coefficients of the relation (12) were chosen in order to work with integer and positive values of the abscissa and to get a sufficient set of sampled values on the horizontal axis.

Figure 4 illustrates the dispersion functions that we obtained for various void ratios ranging from Rv=0.33 for R=16 to Rv=0.02 for R=1, where R is the radius of billiard balls expressed in pixels.

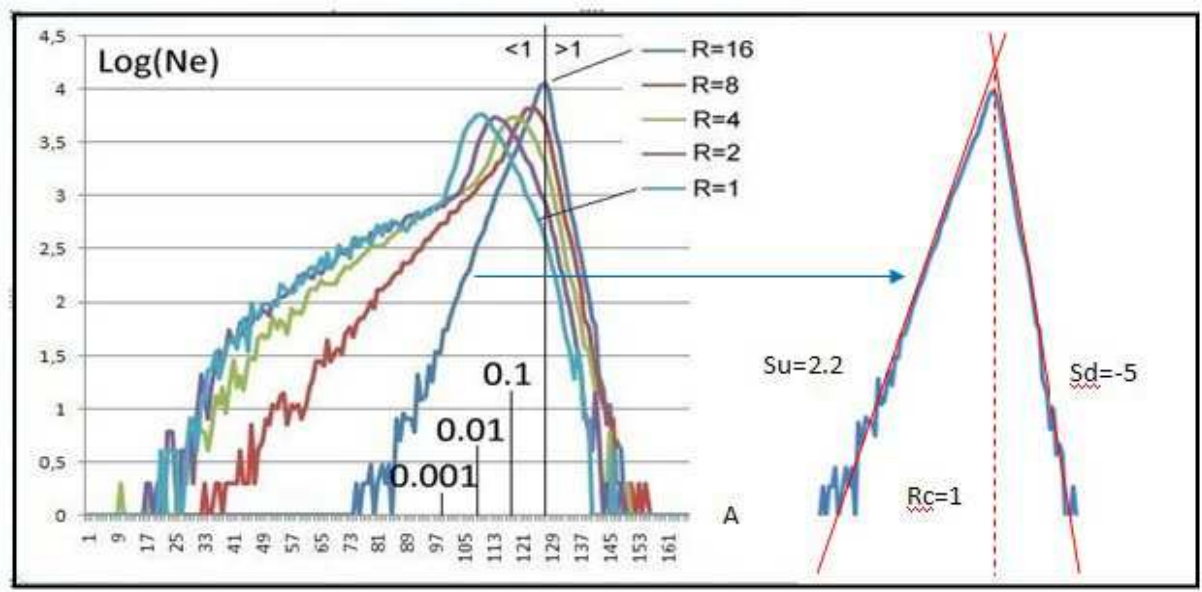

Fig. 4. Distribution of the variation of velocity differences (velocity dispersion) calculated for various void ratios obtained by decreasing the radius R of billiard balls from 16 to 1. Note the important increase in differences of velocity when the void ratio augments: when R varies from 16 to 1 the average value of the distribution varies from 0.1 to 0.001.

Figure 4 shows that when the void ratio decreases from 0.33 to 0.01, the average ratio of velocity differences, after a shock, increases from about 10 to 1000. This can be explained by the fact that decreasing the void ratio increases ball curvature and thus velocity dispersion. We will develop some fundamental consequences of this point during the discussion.

We also note that for the higher void ratio (0.33 for *R*=16) the dispersion function takes a triangular shape with two asymmetric slopes to which we estimated the slope values of *Su*=2.2 and *Sd*=-5. The top of this triangle is the most probable value of *An*, approximately corresponding to a dispersion factor Rc=1, meaning no dispersion at all. Note that the slope differences cause A to be less than 1 most of the time, meaning that shocks are mostly dispersive.

We have exploited the characteristics of this triangle curve to calculate the evolution of *Nc* for higher values of *Nb* (until $2^{17}$) and lower values of $\varepsilon_p$ (until $2^{-45}$). To valid this calculation we first verified that the dispersion distribution remained invariant in respect to *Nb* as long as the void ratio remained constant, and we also verified this invariance when *ΔP* increases from its lower scale (*ΔP* ~ $2^{-35}$) to the higher one (*ΔP* ~1 ). This invariance can be explained by the fact that the higher value of *ΔP* is low enough compared to the radius of the billiard balls (*R*=16).

We used this triangular distribution to extrapolate our void ratio 0.33 simulation until $2^{17}$ ~ 130.000 balls by taking an initial velocity difference equal to $\varepsilon_p$ (as $\varepsilon_p$ is also quantifying velocity for adimensional time). We then successively multiplied the difference by different ratios randomly chosen in respect to their distribution. We stopped these operations when the critical step was reached (difference > 1 pixel) to collect the *Nc* value. We repeated this process as long as necessary to collect a statistically robust *Nc* average.

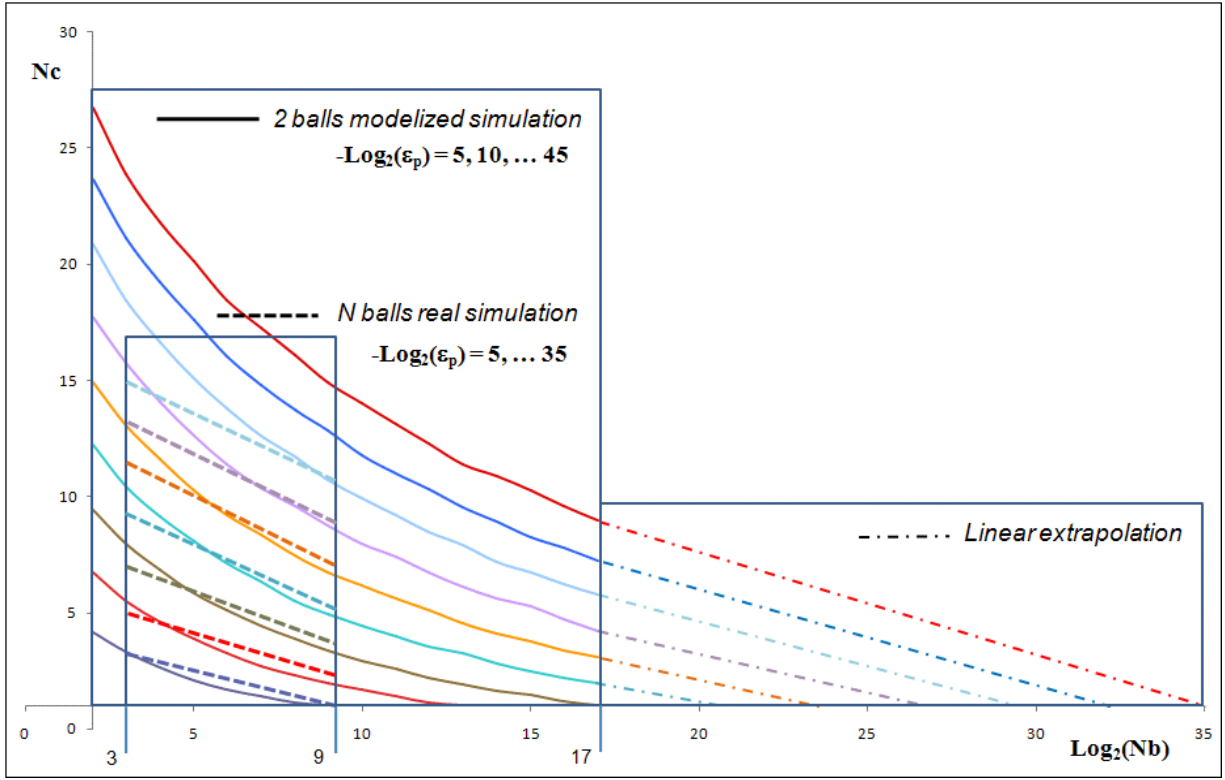

Fig. 5. *Nc* variation curves versus $Log_2(Nb)$ et $-Log_2(\varepsilon_p)$ for the void ratio $Rv=0.33$ : Unbroken lines represent a simulation using a two-ball statistical model (with triangular velocity dispersion function). Dotted lines on the left represent the real simulation using all shock—calculations. Dotted lines on the right represent an extrapolation for making an approximate estimate of the *Nb* values for which the axis *Nc*=1 is reached.

We calculate the *Nc* average for 9 values of $\varepsilon_p$ ( $2^{-5}$ to $2^{-45}$ ) and 16 values of *Nb* ($2^3$ to $2^{17}$ ). The figure 5 shows the global result of our study, where on the left we added dotted lines representing the result of our precedent study and on the right a hypothetical linear extrapolation for $Nb > 2^{17}$ that we will discuss further. We note a rough correspondence between the results of the two different simulations, though the decrease with $Log_2(Nb)$ tends in both cases to produce similar slopes for higher values of *Nb*. The main differences between these results are observed for high values of *Nc* and low values of $\varepsilon_p$.

These differences can be explained by the limited computing precision of 64-bit which introduces arbitrary information at the lowest resolution. Such an error, negligible at small *Nc*, increases however with the number of shocks and is then propagated to higher level bits until the $\varepsilon_p$ resolution is reached, thus the calculations begin to be biased. So the lower the $\varepsilon_p$ value, the sooner the bias begins and keeps on increasing with the average number of shocks *Nc*. The consequence is that we have to consider the two-balls model as more reliable than the real simulation because only the latter is affected by this bias by calculating all the shocks.

Now the following question is to evaluate the possibility for extrapolating these results so as to find out the asymptotic behavior of *Nc* curves when *Nb* → ∞. We have two objective reasons for arguing whatever $\varepsilon_p$, all *Nc* curves bisect the axis *Nc*=1 instead of tending towards the *Nc*=1 value. The first reason is that we can observe that the first three curves calculated for $\varepsilon_p = 2^{-5}$, $\varepsilon_p = 2^{-10}$ and $\varepsilon_p = 2^{-15}$ respectively bisect the axis *Nc*=1 when $Log_2(Nb)$ is approximately equal to 9, 12.5 and 16 (figure 5). The second is that we can show on Figure 6 that whatever $\varepsilon_p$, it is always possible to find a value of *Nb* for which the probability for two trajectories to diverge after only one shock is around 100%, so that the result is *Nc*=1.

Figure 6 illustrates the probability distributions of *Nc* when $\varepsilon_p$ varies from $2^{-5}$ to $2^{-45}$. Note that the distribution curves are approximately parallel when *Nc*→1, so that they always bisect the vertical axis *Nc*=1. This property, which means that the divergence probability is never equal to zero, whatever *Nc*, can be explained by the existence of a borderline case in the behavior of coupled balls: a shock occurs for one of them but not for the other. Figure 6 shows this borderline case for two balls with the same velocity and two perpendicular trajectories, so that they either collide or just graze past each other when their center axis is at 45°. Observe that in the case of collision or grazing, the velocity values are totally different whatever the $\varepsilon_p$ *value*, which explains why a divergence always remains possible for *Nc*=1.

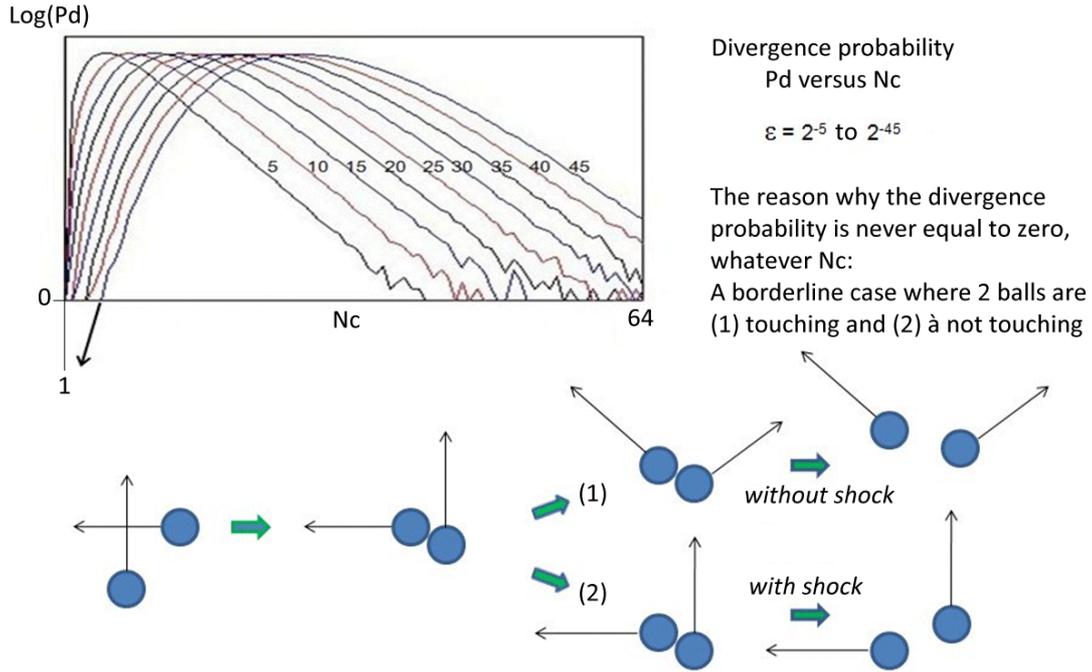

Fig. 6. Top left, logarithm distributions of the probability *Pd* of divergence after a number of shocks *Nc*, in function of $\varepsilon_p$. At the bottom, illustration of the borderline case where two balls are just touching which explains why *Pd* is never equal to 0, whatever the *Nc* value.

We conclude that all the curves plotted on Figure 5 bisect the axis Nc=1, meaning that to adopt a linear model to express the decrease of Nc when Nb tends to infinity is well justified:

$$Nc \sim A + B\,Log(\varepsilon) + C\,Log(Nb) \tag{13}$$

A, B and C are approximately constant values in restrictive intervals of Log(*Nc*) and Log(*ε*). For example, if we suppose that Log(*Nb*)>10 and Log(*ε*)< 50, we can determinate the A,B and C values:

$$Nc \sim 2.8 + 0.21\,Pa - 0.35\,Log(Nb) \tag{14}$$

Where Pa = (-Log(ε)) = (Pi – Pc) is the additional information relative to the precision corresponding to the figures after the decimal point of the phase coordinates. We can then find that when we increment *Pa* and then double the precision, we only have to multiply the number of balls by 1.5 to keep the same critical instant for which the billiard becomes indeterminist.

This means that for a given *Nc*, on which depends the maximum of reliable information (*Nb x Nc x Pc*) that is possible to calculate for a given ball, *Pc* being the precision required for calculation, the additional information that needs to be included in its initial conditions is also proportional to the total number of balls. We can then reformulate *Ib* (10) by using an adimensional time N corresponding to the average number of shocks per ball and also by using *Pc* and *Pa*, for which we have (*Pa + Pc*) < 64 bits:

$$Ib = Nb\, N\, (Pa\, (1 - N/Nc) + Pc) \qquad \text{if } 0 < N < Nc \qquad (15)$$

Thus, the formula (15) expresses the fact that the loss of the information of the billiard ball is linear, which can be explained by a Lyapunov dispersion relative to each shock and to the fact that we consider here only the global averaged statistical effect. We will recall that the residual information *lb = Nb x Nc x Pc* is also lost when *N > Nc* and is lost even more quickly, due to the contagion phenomenon, but we cannot quantify this loss after the critical step because we stop calculations.

It appears from (14) and (15) that for only one phase coordinate the information loss is equal to *Pi/Nc* and increases with Log(*Nb*) for a given *Nc* value. For the values of $\varepsilon_p$ that we sampled and for low values of *Nb*, this loss is around 3 bits per coordinate. Figure 7 illustrates the variation of this loss for *Pa* = 30 bits and *Pc* = 10 bits, so *Pi* = 40 bits: note that when *Nb* increases from 200 to 5000 the information loss grows from 3 to 5 bits per shock and per coordinate (30/10 to 30/6). On this figure, the information relative to initial conditions is represented in green and the information relative to the shock calculations is represented in blue. Each rectangle corresponds to 10 bits, so that we can easily calculate the information contained in one ball trajectory of Nc successive coordinates at the Pc precision, which varies from 100 bits to 60 bits as *Nc* decreases from 10 to 6.

Now if we compare the quantity of information contained in the calculated trajectories with that contained into the initial conditions, we observe the following paradox: this latter quantity can sometimes be higher than the calculated one. This is explained on the one hand by the fact that the initial precision can greatly exceed the one required (40 bits here, instead of 10 bits) and on the other by the fact that in order to memorize the trajectory of one ball, we simply need to memorize successive positions because velocities can be deduced from positions. This is not the case for initial conditions for which we need both positions and velocities. In Figure 7 we have shown the borderline case for which the paradox appears: for *Nb* = 1000 balls and then *Nc* = 8 the trajectory information, equal to 8 x 10 = 80 bits, is exactly equal to the initial condition information which is 2 x 40 = 80 bits. As it is difficult to express this paradox in a few words, we decided to call it "the demon of determinism".

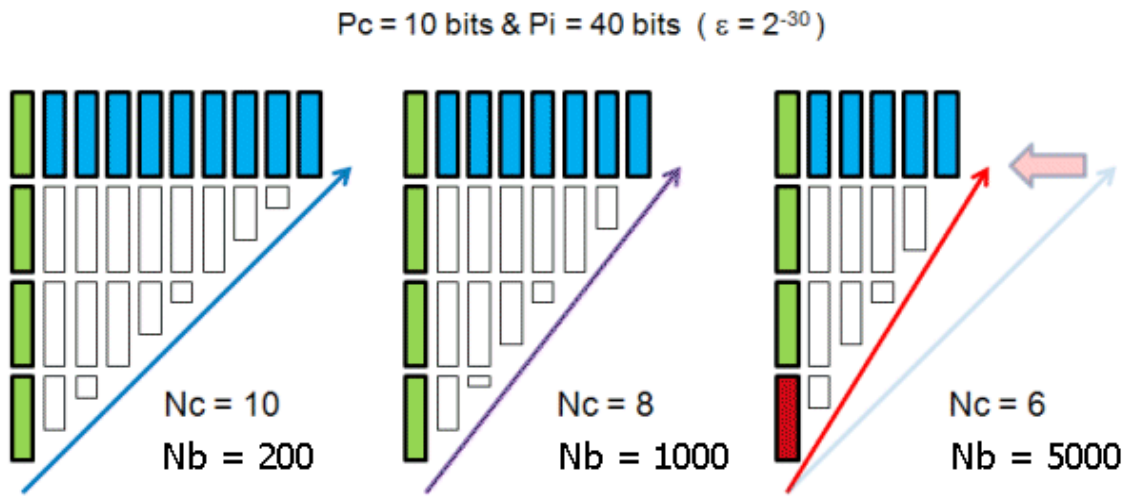

Fig. 7. Illustration of the emergence of the "demon of determinism paradox" for one phase coordinate when $Nb$ increases. Colored rectangles are 10-bit blocks. Green blocks correspond to the information required for initial conditions whose precision is $Pi$=40 bits. Blue blocks correspond to the calculated information whose precision is $Pc$=10 bits. Note that when $Nb$=1000 the initial information is equal to the calculated information. Red blocks correspond to the excess information that occurs when the calculated value is lower than the initial one.

The condition for this paradox to emerge is the following:

$$Pc\ Nc\ Nb < 2\ Pi\ Nb \quad \Rightarrow \quad Nc < 2\ Pi\ /\ Pc \tag{16}$$

As $N$ tends towards 0 when $Nb$ tends towards infinity, this inequality is systematically satisfied above a certain value of $Nb$. So, we conclude whatever the precisions $Pi$ and $Pc$, there always exists a $Nb$ threshold above which the billiard becomes indeterminist. Figure 8 illustrates this generality since it shows the limit points for the paradox to be realized. For a fixed value of $Pc$ and different increasing values of $Pi$, the limit points are approximately aligned along a straight line whose slope is always positive, because $Nc$ necessarily increases with $Pi$. It is also interesting to note the diminution of $Nc$ when we increase $Pc$ while $Pi$ remains constant.

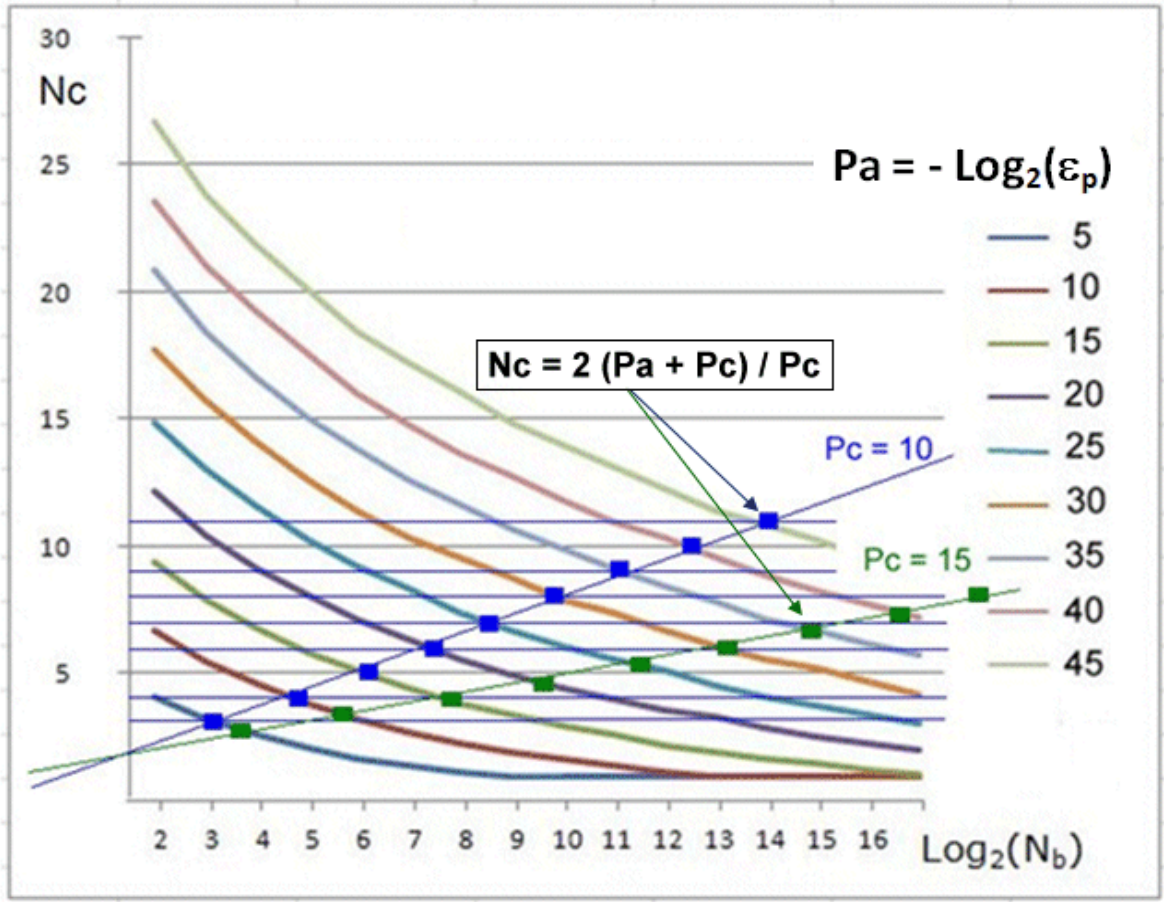

Fig. 8. Illustration of information paradox generality and its emergence conditions: square points correspond to *Nc* and *Nb* values for which initial information is equal to calculated information: when *Pi* is increasing, there is always a *Nb* value for which this limit situation occurs and for a given *Pi*, the effect of increasing *Pc* is to increase the *Nb* limit value, but this also reduces the duration of valid determinist calculations (number of shocks *Nc*)

To conclude this numerical study, we wish to focus on two points:

First, whatever the size of the elementary space quantum, however small it is, the loss of phase information of a billiard game becomes absolute above a certain time or average number of shocks per ball, rendering the game indeterminist or behaving as if the balls were quantum particles. This indetermination resulting from chaos is well known, but we have pointed out that the deterministic time decreases as the number of balls increases.

The second point that we call "the demon of determinism" seems to us more important, because the validity of deterministic calculations (with native or statistical equations) is often considered as depending only on a sufficient precision of initial conditions. But we have shown that whatever this precision, however small is it, there is always a critical number of balls for which the information that is necessary to memorize initial conditions is superior to the information corresponding to valid calculations. This point seems critical to us because it appears to raise a fundamental problem about the validity of information coming from equations, whether they are native or statistic, even in a continuous space.

## III. DISCUSSION

The concept of physical information defined by (2) is involving an epistemic status of uncertainty in classical physics which is no more a lack of information of the observer, but a lack of information of the universe itself. Within a deterministic framework, uncertainty is generally considered the result of the unpredictability inherent to a limit of precision of initial conditions. In this paper, we interpret uncertainty as linked to physical information and no more to observer information, because it is the consequence of our hypothesis that density of physical information is a finite quantity. This is easily conceivable into a discrete universe, but more difficult to understand into a continuous one.

Our results question the implicit idea that the determinism of a system could be guaranteed by unlimited precision of initial conditions, even into a continuous space. Indeed, we have raised a paradox that we qualify as a "demon of determinism" inasmuch as it expresses a strange situation apparently inherent to certain chaotic or dispersive systems: the valid information that can be extracted from a predictive model that calculates their evolution through time has a maximum that can be much lower than the entire amount of information contained in the initial conditions.

A way of by-passing this paradox is to denounce a false one, by arguing that the information calculated by our determinist model is really superior to the initial conditions information, but that a part of it spreads out at levels of resolution lower than that required. In Figure 9, this part corresponds to the white rectangles which complete the triangle delimited with green and blue 10-bit rectangles.

However, not only could this overly low resolution data not be required, if for example non observable, but it could also have no sense at all, since according to our hypothesis the maximum information contained in any space volume is finite and could thus limit the initial information. Another delicate point is that even if space is continuous and phase information is an unlimited quantity, the fact that the calculable and valid information can be much lower than the information that has to be introduced for initial conditions, raises a fundamental problem. Any predictive model should indeed be able to provide a calculation which plays the role of a data compression algorithm. In particular, it should be able to compress the data relative to the trajectories of balls in a billiard into a set of initial conditions occupying much less memory, yet we observe the opposite.

We then qualified this paradox as a "demon of determinism" because it raises a problem which is much more problematic than the well known unpredictability of chaotic systems, which let us think that the determinism of a system would be guaranteed by unlimited precision about the initial conditions. If we consider this unpredictability from the angle of physical information, we observe a crazy situation, in the sense that observable information in a universe of information could be infinitely low compared to the physical information that would be necessary to assure its existence.

To avoid this awkward situation while keeping a physical sense to information, the solution that imposes itself consists of respecting our hypothesis of a discrete space and physical information whose density is limited in every space volume. Now we have shown in the introduction that this hypothesis leads as a direct consequence to the Heisenberg principle. This serves to legitimize our interpretation, which consists in stating that the indeterminism that emerges in a billiard after the critical instant is actually quantum in nature.

Our observations relative to the high increase in dispersion (figure 4) when we increase the void ratio or decrease the balls radius acts to support this interpretation. The smaller the radius of the balls, the more important the loss of information caused by a shock. If we could generalize the validity of our results to particles whose interactions are no longer elastic but electromagnetic, then the quantum indeterminism would itself appear to be a direct consequence of the space geometry. However, we will not focus on this subject in the present paper.

Whatever the origin of quantum states, our hypothesis - that physical information has a finite density - can explain the difference between quantum and classic behavior: any system becomes a quantum system when it loses its phase information, this loss being a phenomenon that characterizes highly dispersive systems such as those containing numerous objects that interact with each other. So the information lost by a system has to be considered as a really lack in the purely classical physical reality, until a compensatory mechanism introduces information again, allowing the system to restore classic behavior without superposed trajectories. In quantum physics, this mechanism involves the two aspects of decoherence and observation.

Decoherence mechanism is due to the interaction between a quantum sub-system and its environment. This environment is usually the classical reality of a laboratory or an experiment that can be considered as a non quantum system enclosing the quantum sub-system. The result of their interaction is that the sub-system cannot maintain quantum states that are not coherent with the well defined states of the enclosing system, thus resulting in a reduction of the superposed states of the sub-system: this can be considered as a transmission of physical information from environment to a quantum sub-system that therefore becomes a classical system, at least temporarily.

The observation mechanism is a direct way of informing quantum reality, resulting in a collapse of the quantum wave of the measured particle that is very well known and even the center of an epistemological debate which we avoid here, the problem being that the source of information that informs the reduced states is unknown: Quantum randomness? Hidden variables? Extra dimensions [24]? According to Antoine Suarez, this information is a quantum randomness that could be controlled from outside space-time by free will [25]. Whatever its source, we find interesting to propose that observations could be a way to transmit information from this unknown source to everything that is observed, which is the case for environment itself. The major part of information transfer would then be due to the process of decoherence, initiated by already informed parts of the universe (that don't lose their information). This could explain why reality always appears to us as purely classical, knowing that only the parts that are simultaneously non observable, dispersive and completely isolated could maintain quantum states.

With such a proposition, the irreversibility of quantum measurement would have the same origin as the irreversibility in classical physics: a loss of information during interactions, which quantum randomness introduced by observation is not able to recover as before. This loss would be all the more fast as the dimension of objects (particles, balls, molecules…) would be small. So irreversibility appears to be directly connected to our fundamental hypothesis that the density of physical information is finite and limited everywhere. Note that such an hypothesis is essential if we consider to live into a "cyberspace" of information where anything is the result of a "bit with bit" calculation, like in computers.

Our model of physical information is not without impact on our habitual conception of reality, because it calls to consider our observable reality as a "cyberspace" of information. A minuscule part of this reality would be informed directly by observations and the major part by a cascading decoherence processus. It would subsist only quantum systems which are not informed because they are completely isolated or they lose their information too quickly.

In particular, it could be the case of isolated gas, where mixing interactions between molecules are responsible for increasing entropy.

In our model, entropy and information are equivalent concepts and opposite quantities, and so the increase in entropy during the mixing process corresponds to a loss of information or to quantum behavior of gas molecules. This explains the indiscernability of molecules that is only admitted today as a principle and it offers a clear solution to the Gibbs paradox by confirming the Pauli hypothesis of random phases [21] in statistical physics.

From this point of view, therefore, it is interesting to reconsider the interpretation of Brillouin [22] and Szillard [1] about Maxwell's Demon: they exorcise it by saying that any observation of a molecule introduces information into the system that decreases its entropy with an energetic cost due to the measurement at least equal to kTln(2), although the physical sense of this information does not appear clearly in this interpretation, which maintains a subjective character. This subjectivity disappears if the molecules display quantum behavior, because the information brought to them regains a physical sense which is the phase state reduction through quantum measurement.

We can thus summarize the physical sense of information established by our model in two points:

(1) The information lost during multiple interactions such as mixing is really lost by the physical system through a dispersive mechanism that generates quantum phase states, thus explaining its irreversibility.

(2) The physical information gained by a quantum system which becomes at least partially a classical one, is gained directly by means of an observation or measurement or indirectly by means of the decoherence process during which the exterior system transmits physical information.

The difference we make between directly or indirectly acquired information comes from the fact that in the first case of observation, the source is outside the universe of classical information, taking into account the properties of quantum wave function collapse: the information acquired in a reduction of phase states does not exist before the measurement or it should depend on non local hidden variables [23]. This is not the case with the process of decoherence that involves only information already contained into the local and classical environment of the system.

Observation thus appears to involve a mechanism that is still poorly understood but able to introduce new information into a computable universe, meaning that this information is not already included in this classical universe. Therefore we conceive this universe as a sub-universe immersed in a quantum global universe, one containing the source of information that could be transmitted by observations. In a certain measure that remains unknown, this could compensate for the natural loss of information that affects the computable universe.

## IV. CONCLUSION

The heart of this article is an asymptotic study of statistical results of digital simulations of a billiard proposed as a simple model for a better understanding of the mechanisms of loss of information or increase of entropy in a gas. Though purely technical, it has the fundamental advantage of highlighting a paradox we call "the demon of determinism", something that causes problems when we try to find a physical sense for phase information: not only should we accept having to deal with calculation models that may consume more information than this we are trying to calculate, but also, in order to preserve absolute determinism physical information should be infinite in closed parts of space. Now if we consider that physical information is really quantified by a quantum $k\ln(2)$ in relation to entropy, this conclusion appears unacceptable.

We then decided to postulate that the density of physical information should be a finite quantity everywhere, by giving it a purely classical sense, something that had the advantage of being simpler and more intuitive than algorithmic complexity. What is notable here is that the Heisenberg principle can be interpreted as a direct consequence of these hypotheses. This then leads us to suggest a transition from classical to quantum states in order to understand the loss of information and thus send the question of determinism back to quantum mechanics. It then becomes interesting to note that this suggests quantum mechanics could be a natural extension of classical mechanics, not at all incompatible or conflicting, since purely classical models seem to fail to give a physical and objective sense to phase information in a space that is not quantified.

By using our physical information model to clarify the famous thermodynamic ambiguities raised by Gibbs and Maxwell, we could verify its pertinence and then conclude that a classical dispersive system could lose its physical phase information when it is isolated from any process of quantum reduction by decoherence or direct observation. Any phase state could then become at least partially a quantum one and again recover its physical information when it interacts with an "informed" environment or a measuring device. However, where this information is directly brought about by observation, according to quantum mechanics it could be not already included in our classical universe of physical information. We could then claim that observation could add physical information to the universe.

We may well wonder what *observation* and its mechanism actually are, as well as the information source that could inform our classical reality through observation, but these questions are already part of the debate in quantum mechanics. Our own sentiment is simply that our classical reality could be immersed in a more global quantum world where notions of space, time and causality could be upset. Without speculating further, it remains important to notice that this interface function of observation could question the second law of thermodynamics, according to which the entropy of the global universe should only increase. Which global universe? If we consider our universe of classical information and the physical sense of entropy that emerges from our model to be a measurement of quantum indetermination, it turns out that observation and then consciousness could compensate for increase in entropy, thus explaining why it could decrease or remain stable in living systems. However, we have no idea at all about the extent to which it could happen, and it is possible that this contribution could be minuscule and even negligible, as everything leads us to believe that the classical world around us is already perfectly informed.